\documentclass[dvips]{plain}
\usepackage{supertabular,lscape,epsfig}
\usepackage{amssymb}
\usepackage{amsmath}
\usepackage{wasysym}

\SetPages{1}{0}

\SetVol{0}{2020}

\begin{document}

\begin{Titlepage}
\Title{Search for planets in hot Jupiter systems with multi-sector TESS photometry. I. No companions in planetary systems KELT-18, KELT-23, KELT-24, Qatar-8, WASP-62, WASP-100, WASP-119, and WASP-126.}
\Author{Gracjan~Maciejewski}
{Institute of Astronomy, Faculty of Physics, Astronomy and Informatics,
         Nicolaus Copernicus University, Grudziadzka 5, 87-100 Toru\'n, Poland,
         e-mail: gmac@umk.pl}

\Received{October 2020}
\end{Titlepage}

\Abstract{Origins of giant planets on tight orbits, so called hot Jupiters, are a long-lasting question in the planetary formation and evolution theory. The answer seems to be hidden in architectures of those systems that remain only partially understood. Using multi-sector time-series photometry from the Transiting Exoplanet Survey Satellite, we searched for additional planets in the KELT-18, KELT-23, KELT-24, Qatar-8, WASP-62, WASP-100, WASP-119, and WASP-126 planetary systems using both the transit technique and transit timing method. Our homogenous analysis has eliminated the presence of transiting companions down to the terrestrial-size regime in the KELT-23 and WASP-62 systems, and down to mini-Neptunes or Neptunes in the remaining ones. Transit timing analysis has revealed no sign of either long-term trends or periodic perturbations for all the studied hot Jupiters, including the WASP-126~b for which deviations from a Keplerian model were claimed in the literature. The loneliness of the planets of the sample speaks in favour of the high-eccentricity migration mechanism that probably brought them to their tight orbits observed nowadays. As a byproduct of our study, the transit light curve parameters were redetermined with a substantial improvement of the precision for 6 systems. For KELT-24~b, a joint analysis allowed us to place a tighter constraint on its orbital eccentricity.}{stars: individual: KELT-18, KELT-23, KELT-24, Qatar-8, WASP-62, WASP-100, WASP-119, WASP-126 -- planets and satellites: individual: KELT-18 b, KELT-23 b, KELT-24 b, Qatar-8 b, WASP-62 b, WASP-100 b, WASP-119 b, WASP-126 b}


\section{Introduction}

Hot Jupiters are arbitrarily defined as planetary gas giants orbiting their host stars within less than approximately $10$ days (Udry \& Santos 2007) and having masses above 0.1 -- 0.2 $M_{\rm Jup}$ and radii greater than approximately $0.4$ $R_{\rm Jup}$ (Winn \& Fabrycky 2015), where $M_{\rm Jup}$ and $R_{\rm Jup}$ are the mass and radius of Jupiter respectively. Their origin has remained a subject of debate since the discovery of 51 Pegasi b -- the first exoplanet around a Sun-like star (Mayor \& Queloz 1995). Assuming that giant planets are formed beyond their host star's water frost line, hot Jupiters must have migrated to their tight orbits observed nowadays. 

In the disk migration scenario, a giant planet spirals into its host star due to tidal interactions within a protoplanetary disk in early stages of system evolution, ultimately reaching an inner edge of the disk as a hot Jupiter (Lin \& Papaloizou 1986, Lin \etal 1996). Although the numerical simulations show that such a passage may trigger formation of both inner and outer low-mass planetary companions (Mandell \etal 2006), the statistical studies show that hot Jupiters are rather alone (Wright \etal 2009, Latham \etal 2011, Steffen \etal 2012) or accompanied by massive planets on wide and eccentric orbits (\eg Bonomo \etal 2017). So far, just a handful of hot Jupiters in compact systems has been found: WASP-47~b with a Neptune-sized outer planet and a super-Earth inner companion (Becker \etal 2015) and accompanied by a distant massive planet (Neveu-VanMalle \etal 2016), Kepler-730~b with an interior Earth-sized planet (Ca{\~n}as \etal 2019), TOI-1130~b with a Neptune-sized inner companion (Huang \etal 2020), and WASP-148~b being perturbed by a more massive outer companion near the 4:1 mean-motion resonance (H\'ebrard \etal 2020). 

Population analysis suggests that only $\approx15$\% of hot Jupiters could have undergone disk migration (Nelson \etal 2017). The remaining vast majority of those planets must have arrived at their tight orbits via high-eccentricity migration. In this mechanism, an orbital eccentricity of a gas giant is excited by close encounters with other planet (Rasio \& Ford 1996) or by Kozai's interaction with a distant binary companion (Eggleton \& Kiseleva-Eggleton 2001). If a periastron distance is small enough the orbit is being circularised and shrinks due to star-planet tidal dissipation. Such dynamical instability is expected to be destructive for low-massive bodies in the system (Mustill \etal 2015), inevitably leaving hot Jupiters in seclusion.

Learning the architecture of the planetary systems with hot Jupiters helps to build up a complete picture of planetary formation and evolution. Because of their physical properties, hot Jupiters are relatively easy to detect from the ground using either the Doppler technique or photometric transit method. This feature leads to inevitable observational biases (Kipping \& Sandford 2016). Space-borne facilities, such as Kepler (Borucki \etal 1997) and CoRoT (Rouan \etal 1998), allowed us to discover the wealth and diversity of smaller exoplanets. This was achieved by providing uninterrupted photometric time series of a high quality and short cadence. Low-mass planetary companions of hot Jupiters can be detected directly with the transit method or inferred from transit timing variations (TTVs). The latter technique is based on time shifts in transit times of a known planet which are a manifestation of gravitational perturbations induced by the other body. This method is particularly effective in configurations close to mean motion resonances, and can be applied to planets that remain inaccessible for other detection methods (\eg Ballard \etal 2011).
 
The Transiting Exoplanet Survey Satellite (TESS, Ricker \etal 2014) surveys the brightest stars for transiting exoplanets with a 85\% sky coverage. Each sky sector is continuously observed for almost 4 weeks. As sectors overlap for higher ecliptic latitudes, photometric time series spanning even 1 year are acquired. In our project, we analyse the TESS photometry for a sample of planetary systems with transiting hot Jupiters that were observed in at least 3 sectors. This multi-sector photometry not only allows us to probe both short- and long-term TTVs, but also increases chances for transit detection of additional planets on wide orbits. In this paper, we present the methodology used and explore 8 systems: KELT-18 (McLeod \etal 2017), KELT-23 (Johns \etal 2019), KELT-24 (Rodriguez \etal 2019, Hjorth \etal 2019), Qatar-8 (Alsubai \etal 2019), WASP-62 (Hellier \etal 2012), WASP-100 (Hellier \etal 2014), WASP-119 (Maxted \etal 2016), and WASP-126 (Maxted \etal 2016).    
 

\section{Systems of the sample}

KELT-18~b was discovered by the the Kilodegree Extremely Little Telescope (KELT, McLeod \etal 2017). It has a mass of $\approx 1.2$ $M_{\rm Jup}$ and orbits its F4-type host star within 2.9 d. It was identified as one of the most inflated planets due to its radius of $\approx1.6$ $R_{\rm Jup}$. The system is accompanied by a probably gravitationally bound less-luminous star. 

The G2-type star KELT-23A is a probable member of a long-period binary star system with a low-mass companion. It hosts a $\approx 0.9$ $M_{\rm Jup}$ and $\approx 1.3$ $R_{\rm Jup}$ planet on a 2.3-d orbit (Johns \etal 2019). Future orbital evolution was found to be driven by star-planet tidal interactions. The planet is predicted to spiral into its host with a time scale of gigayears.

The star HD 93148 (= KELT-24 = MASCARA-3) was independently found to host a transiting giant planet by KELT  (Rodriguez \etal 2019) and the Multi-Site All-sky CAmeRA (MASCARA, Hjorth \etal 2019). The planet has a mass of $\approx 5 \, M_{\rm Jup}$ and a radius of  $\approx 1.3 \, R_{\rm Jup}$. The orbital period is $\approx 5.6$ days. Interestingly, the radial-velocity (RV) measurements suggest that the planet's orbit is not circularised. Rodriguez \etal (2019) report the eccentricity $e = 0.077^{+0.024}_{-0.025}$ and Hjorth \etal (2019) bring $e = 0.050^{+0.020}_{-0.017}$.

The planet Qatar-8~b, the orbital period of which is $\approx 3.7$ days, was discovered by the Qatar Exoplanet Survey (Alsubai \etal 2019). With its mass of $\approx 0.4 \, M_{\rm Jup}$ it was classified as a hot Saturn. Its radius was found to be bloated to $\approx 1.3 \, R_{\rm Jup}$.  

The Wide Angle Search for Planets survey (WASP) found that the F7-type star WASP-62 is orbited by a $\approx 0.4 \, M_{\rm Jup}$ planet within $\approx 4.4$ days (Hellier \etal 2012). Both lithium abundance and gyrochronology show that the system is $\approx 0.7$ Gyr old. The planet is bloated to $\approx 1.4 \, R_{\rm Jup}$. The occultation timing shows that the planet's orbit is slightly eccentric with $e \cos \omega= 0.00614 \pm 0.00064$ (Garhart \etal 2020). From an analysis of the Rossiter-McLaughlin (RM) effect, Brown \etal (2017) found that the orbit of WASP-62~b is slightly misaligned with the sky-projected spin-orbit angle $\lambda$ between the orbital plane and the host star's equatorial plane equal to $19.4^{+5.1}_{-4.9}$ degrees. Kilpatrick \etal (2017) analysed a near-IR photometric time series with an occultation of WASP-62~b to find that the planet's atmosphere has a rather moderate albedo and efficient energy recirculation. Further spectroscopic observations have revealed the presence of water vapour and likely iron hydride (Skaf \etal 2020).  

The planet WASP-100~b orbits its F2 host star within $\approx 2.8$ days (Hellier \etal 2014). It has a mass of $\approx 2 \, M_{\rm Jup}$ and inflated radius of $\approx 1.7 \, R_{\rm Jup}$. The orbit was found to be circular (Garhart \etal 2020) but its plane is significantly misaligned with $\lambda = 79^{+19}_{-10}$ degrees (Addison \etal 2018), indicating a nearly polar orbit orientation.

The host star in the WASP-119 system was found to be a 8 Gyr analogue of the Sun (Maxted \etal 2016). It is orbited by a $\approx 1.2 \, M_{\rm Jup}$ and $\approx 1.4 \, R_{\rm Jup}$ planet with a period of $\approx 2.5$ days.

The planet WASP-126~b needs $\approx 3.3$ days to complete an orbit around its G2-type host (Maxted \etal 2016). With a mass of $\approx 0.3 \, M_{\rm Jup}$ and a radius of $\approx 1.0 \, R_{\rm Jup}$, it is a hot analogue of Saturn. Interestingly, Pearson (2019) analysed the TESS photometric data from sectors 1--3 and postulated that the planet exhibits a possible TTV signal with an amplitude of $\approx 1 \, {\rm min}$ and a period of $\approx 25 \, {\rm d}$. Those timing deviations from a Keplerian solution could be induced by a non-transiting $\approx 60 \, M_{\oplus}$ planet on a $\approx 7.7 \, {\rm d}$ orbit.

Observational properties of the examined systems are summarised in Table~1.

\MakeTable{lcccccc}{12.5cm}{Observational properties of the systems of the sample.} 
{\hline
System  & RA (J2000)  & Dec (J2000) & $m$  & Distance & $d_{\rm tr}$        & $\delta_{\rm tr}$ \\
        & hh:mm:ss.s  & $\pm$dd:mm:ss   & (mag)    & (pc)     & (min)                 & (ppth)            \\
\hline 
KELT-18 & 14:26:05.8  & +59:26:39   & 10.16      & $324.4\pm2.5$ & $284.1^{+9.1}_{-8.0}$ & $7.15^{+0.08}_{-0.08}$ \\
KELT-23 & 15:28:35.2  & +66:21:32   & 10.31      & $126.7\pm0.4$ & $136.7^{+1.0}_{-1.0}$ & $17.43^{+0.15}_{-0.15}$ \\
KELT-24 & 10:47:38.4  & +71:39:21   & 8.33       & $96.8\pm0.3$ & $275.4^{+7.9}_{-6.8}$ & $8.11^{+0.06}_{-0.07}$ \\
Qatar-8 & 10:29:39.1  & +70:31:38   & 11.71      & $283.8\pm2.6$ & $250^{+25}_{-20}$ & $10.29^{+0.63}_{-0.53}$ \\
WASP-62 & 05:48:33.6  & --63:59:18   & 10.21      & $176.5\pm0.6$ & $226.4^{+1.5}_{-1.5}$ & $12.22^{+0.06}_{-0.06}$ \\
WASP-100 & 04:35:50.3 & --64:01:37   & 10.80      & $368.3\pm2.8$ & $225.2^{+3.8}_{-4.5}$ & $6.42^{+0.06}_{-0.06}$ \\
WASP-119 & 03:43:44.0 & --65:11:38   & 12.10      & $305.1\pm2.0$ & $172.8^{+2.3}_{-3.8}$ & $12.17^{+0.22}_{-0.18}$ \\
WASP-126 & 04:13:29.7 & --69:13:37   & 11.09      & $217.9\pm0.9$ & $205.2^{+3.0}_{-5.5}$ & $5.95^{+0.10}_{-0.07}$ \\
\hline
\multicolumn{7}{l}{Coordinates taken from the Gaia Data Release 2 (DR2, Gaia Collaboration \etal 2018).}  \\
\multicolumn{7}{l}{$m$ is apparent brightness in a $V$ band from the Tycho-2 catalogue (H{\o}g \etal 2000),}  \\
\multicolumn{7}{l}{and in a $G$ band from DR2 for WASP-119. Distance is calculated on DR2 parallaxes.}  \\
\multicolumn{7}{l}{$d_{\rm tr}$ and $\delta_{\rm tr}$ are transit duration and transit depth refined in this study.}  \\
\multicolumn{7}{l}{ppth stands for parts per thousand of normalised out-of-transit flux.}  \\
}


\section{Observations and data reduction}

\subsection{TESS photometry}

KELT-18 was observed between August 2019 and April 2020 in 4 sectors. In the first 3 runs, exposures were acquired in the long cadence (LC) mode with exposure times of 30 min. Observations in fourth sector were performed in the short cadence (SC) mode with exposure times of 2 min. KELT-23 was observed in the SC mode in 6 sectors spanned between July 2019 and April 2020. KELT-24 and Qatar-8 were observed between July 2019 and February 2020 in 3 sectors. Exposures were acquired in the LC mode. The remaining four targets were observed with SC. WASP-62, WASP-100, and WASP-126 were monitored between July 2018 and July 2019. WASP-62 was observed in 12 sectors, and the remaining two targets were monitored continuously in 13 sectors. Observations of WASP-119 spanned from July 2018 to May 2019 in 6 sectors. The observations are summarised in Table~2.

\MakeTable{cccccc}{12.5cm}{Details on TESS observations used.} 
{\hline
Sect./ & from--to & $N_{\rm tr}$ & Sect./ & from--to & $N_{\rm tr}$ \\
/Mode  &            &           & /Mode  &          &    \\
\hline 
\multicolumn{3}{c}{KELT-18} &  \multicolumn{3}{c}{WASP-100}  \\
15/LC & 2019-Aug-15--2019-Sep-11 & 9 & 1/SC & 2018-Jul-25--2018-Aug-22 & 10 \\
16/LC & 2019-Sep-11--2019-Oct-07 & 8 & 2/SC & 2018-Aug-22--2018-Sep-20 & 10 \\
22/LC & 2020-Feb-18--2020-Mar-18 & 7 & 3/SC & 2018-Sep-20--2018-Oct-18 & 7 \\
23/SC & 2020-Mar-18--2020-Apr-16 & 8 & 4/SC & 2018-Oct-18--2018-Nov-15 & 7 \\
\multicolumn{2}{r}{total:} & 32 & 5/SC & 2018-Nov-15--2018-Dec-11 & 9 \\
\multicolumn{3}{c}{KELT-23} & 6/SC & 2018-Dec-11--2019-Jan-07 & 7 \\
14/SC & 2019-Jul-18--2019-Aug-15 & 12 & 7/SC & 2019-Jan-07--2019-Feb-02 & 8 \\
15/SC & 2019-Aug-15--2019-Sep-11 & 10 & 8/SC & 2019-Feb-02--2019-Feb-28 & 5 \\
16/SC & 2019-Sep-11--2019-Oct-07 & 10 & 9/SC & 2019-Feb-28--2019-Mar-26 & 8 \\
17/SC & 2019-Oct-07--2019-Nov-02 & 10 & 10/SC & 2019-Mar-26--2019-Apr-22 & 8 \\
21/SC & 2020-Jan-21--2020-Feb-18 & 11 & 11/SC & 2019-Apr-22--2019-May-21 & 8 \\
23/SC & 2020-Mar-18--2020-Apr-16 & 11 & 12/SC & 2019-May-21--2019-Jun-19 & 8 \\
\multicolumn{2}{r}{total:} & 64 & 13/SC & 2019-Jun-19--2019-Jul-18 & 9 \\
\multicolumn{3}{c}{KELT-24} & \multicolumn{2}{r}{total:} & 104 \\
14/LC & 2019-Jul-18--2019-Aug-15 & 5 & \multicolumn{3}{c}{WASP-119}  \\
20/LC & 2019-Dec-24--2020-Jan-21 & 5 & 1/SC & 2018-Jul-25--2018-Aug-22 & 11 \\
21/LC & 2020-Jan-21--2020-Feb-18 & 4 & 2/SC & 2018-Aug-22--2018-Sep-20 & 10 \\
\multicolumn{2}{r}{total:} & 14 & 3/SC & 2018-Sep-20--2018-Oct-18 & 8 \\
\multicolumn{3}{c}{Qatar-8} & 4/SC & 2018-Oct-18--2018-Nov-15 & 8 \\
14/LC & 2019-Jul-18--2019-Aug-15 & 7 & 7/SC & 2019-Jan-07--2019-Feb-02 & 10 \\
20/LC & 2019-Dec-24--2020-Jan-21 & 7 & 11/SC & 2019-Apr-22--2019-May-21 & 9 \\
21/LC & 2020-Jan-21--2020-Feb-18 & 7 & \multicolumn{2}{r}{total:} & 56 \\
\multicolumn{2}{r}{total:} & 21 & \multicolumn{3}{c}{WASP-126}  \\
\multicolumn{3}{c}{WASP-62} & 1/SC & 2018-Jul-25--2018-Aug-22 & 8 \\
1/SC & 2018-Jul-25--2018-Aug-22 & 5 & 2/SC & 2018-Aug-22--2018-Sep-20 & 8 \\
2/SC & 2018-Aug-22--2018-Sep-20 & 6 & 3/SC & 2018-Sep-20--2018-Oct-18 & 5 \\
3/SC & 2018-Sep-20--2018-Oct-18 & 4 & 4/SC & 2018-Oct-18--2018-Nov-15 & 7 \\
4/SC & 2018-Oct-18--2018-Nov-15 & 5 & 5/SC & 2018-Nov-15--2018-Dec-11 & 8 \\
6/SC & 2018-Dec-11--2019-Jan-07 & 5 & 6/SC & 2018-Dec-11--2019-Jan-07 & 7 \\
7/SC & 2019-Jan-07--2019-Feb-02 & 6 & 7/SC & 2019-Jan-07--2019-Feb-02 & 8 \\
8/SC & 2019-Feb-02--2019-Feb-28 & 3 & 8/SC & 2019-Feb-02--2019-Feb-28 & 6 \\
9/SC & 2019-Feb-28--2019-Mar-26 & 4 & 9/SC & 2019-Feb-28--2019-Mar-26 & 8 \\
10/SC & 2019-Mar-26--2019-Apr-22 & 5 & 10/SC & 2019-Mar-26--2019-Apr-22 & 7 \\
11/SC & 2019-Apr-22--2019-May-21 & 6 & 11/SC & 2019-Apr-22--2019-May-21 & 7 \\
12/SC & 2019-May-21--2019-Jun-19 & 6 & 12/SC & 2019-May-21--2019-Jun-19 & 7 \\
13/SC & 2019-Jun-19--2019-Jul-18 & 6 & 13/SC & 2019-Jun-19--2019-Jul-18 & 8 \\
\multicolumn{2}{r}{total:} & 61 & \multicolumn{2}{r}{total:} & 94 \\
\hline
\multicolumn{6}{l}{Mode specifies long cadence (LC) or short cadence (SC) photometry.}  \\
\multicolumn{6}{l}{$N_{\rm tr}$ is a number of complete transits qualified for this study.}  \\
}

Fluxes were obtained with the aperture photometry method applied to target pixel files which were centred at a target position. The LC frames, $15 \times 15$ pixel ($5.25' \times 5.25'$) wide, were extracted from full-frame images with the TESSCut\footnote{https://mast.stsci.edu/tesscut/} tool (Brasseur \etal 2019). The $11 \times 11$ pixel ($3.85' \times 3.85'$) wide SC frames, which are available for pre-selected targets, were downloaded from the exo.MAST portal\footnote{https://exo.mast.stsci.edu}. The light curves were obtained with standard procedures available in the Lightkurve v1.9 package (Lightkurve Collaboration 2018). The aperture mask was manually optimised for each target and each sector to account for blends and inter-sector changes of a stellar profile and orientation of a field of view. The sky background was mapped using the standard-deviation-based thresholding method. A light curve with total source counts was verified by eye and, if needed, a 5--10 sigma clipping was applied in order to remove outlying measurements. Finally, the light curve was normalised by removal of the low frequency trends, which could be caused by stellar variation or systematic effects on time scales much longer than the expected transit duration, using the Savitzky-Golay filter. In that step, data points at transit and occultation phase were masked out using a preliminary transit ephemeris. A filter's window width was optimised to produce the lowest data-point scatter and set to 10 hours in a result of a series of trials.

Some portion of data was lost due to scattered light and downlink breaks. The former effect manifested as an increase of data-point scatter and time-correlated flux ramps. Such corrupted measurements were identified by visual inspection and then cut out.

\subsection{Ground-based photometry for KELT-18}

A transit of KELT-18~b was observed on 2017 February 27 with a 0.6 m and f/12.5 Cassegrain telescope located at the astronomical observatory of the Institute of Astronomy of the Nicolaus Copernicus University (Piwnice near Toru\'n, Poland). The observing run was occasionally interrupted by passing clouds. An SBIG STL-1001E camera, used as a detector, provided a $12' \times 12'$ field of view. Exposures were 12 s long with cadence of 15 s. To increase a signal-to-noise ratio, the time series was acquired through a long-pass filter which blocked wavelengths shorter than 500 nm. The instrument was manually guided to compensate tracking inaccuracies and keep the target star at the same $xy$ position on the CCD matrix. 

A standard data calibration and reduction procedure was performed with AstroImageJ (Collins \etal 2017). The science frames were subjected to dark current and flat field correction. Fluxes were obtained using the aperture photometry method with an aperture size and collection of comparison stars optimised to achieve the lowest data-point scatter. Fluxes were de-trended against time, position on the matrix, and width of a stellar profile, and then normalised to unity outside the transit. The time stamps were converted to barycentric Julian dates in barycentric dynamical time $(\rm{BJD_{TDB}})$ with a built-in converter.  


\section{Data analysis and results}

\subsection{Refining the orbital eccentricity for KELT-24~b}

The orbital elements of KELT-24~b such as the eccentricity $e$, argument of periastron $\omega$, and radial velocity (RV) amplitude $K$ can be refined by a joint reanalysis of Doppler measurements taken from Rodriguez \etal (2019) and Hjorth \etal (2019). The first two of those parameters are particularly important because they are included in transit light curve modeling (Sect.~4.3). In our reanalysis, we used 19 out-of-transit RV measurements from Rodriguez \etal (2019), which were acquired with the Tillinghast Reflector Echelle Spectrograph (TRES; F\H{u}r\'esz 2008) in March and April 2019, and 65 RV observations from Hjorth \etal (2019), secured between April 2018 and May 2019 by the Stellar Observation Network Group (SONG, Andersen \etal 2014). In-transit observations were skipped because our analysis procedure does not account for the RM effect. 

A Keplerian model of the system was constructed with the Systemic 2.18 software (Meschiari \etal 2009). The RV datasets were enhanced with transit times (Sect.~4.4). The orbital period $P_{\rm orb}$, $e$, $\omega$, and $K$ were kept free. To account for instrumental systemics in the absolute RV calibration, a shift between TRES and SONG datasets was allowed to vary. The best-fitting solution was found with the Levenberg-Marquardt method. The parameter uncertainties were estimated as median absolute deviations using a bootstrap run of $10^5$ trials. The best-fitting parameters, together with the values from the literature, are given in Table~3. Our analysis places a tighter constraint on the orbital eccentricity which was found to be closer to 0. Consequently, $\omega$ was found to be constrained weaker. The value of $K$ was found to be consistent within 0.8$\sigma$ with the value reported by Hjorth \etal (2019), and smaller by 1.8$\sigma$ from the result of Rodriguez \etal (2019).

\MakeTable{ l c c c }{12.5cm}{Literature and refined orbital parameters for KELT-24~b.}
{\hline
Parameter         & Rodriguez \etal (2019) & Hjorth \etal (2019) & this paper\\
\hline
$e$               & $0.077^{+0.024}_{-0.025}$ & $0.050^{+0.020}_{-0.017}$ & $0.033\pm0.012$ \\
$\omega$ ($^{\circ}$) & $55^{+13}_{-15}$ & $27^{+38}_{-27}$ & $4 \pm 44$ \\
$K$ (${\rm m}\,{\rm s}^{-1}$) & $462^{+16}_{-15}$ & $415 \pm 13$ & $428 \pm 11$ \\
\hline
}

\subsection{On the orbital eccentricity of WASP-62~b}

Garhart \etal (2020) have found that the occultations of WASP-62~b occur about 23 minutes later than phase 0.5\footnote{Phase 0.0 refers to a transit mid-point.}. Following the procedure we applied to KELT-24~b  (Sect.~4.1), we reanalysed the Doppler and transit and occultation timing data sets in order to place tighter constraints on the orbital parameters. We used 24 out-of-transit RV measurements from Hellier \etal (2012) which were obtained with the CORALIE spectrograph on the Euler 1.2-m telescope at La Silla (Chile) between March 2011 and April 2012. In addition, 8 RV measurements were taken from Brown \etal (2017). They were acquired with the high resolution \'echelle spectrograph HARPS at the 3.6 m ESO telescope at La Silla on 2012 October 12. The mid-transit times (Sect.~4.4) were enhanced with 2 occultation times from Garhart \etal (2020).  

Analysis of posterior distributions reveals that the value of $e$ is weakly constrained by current data sets and is strongly correlated with $\omega$. The values of $\omega$ were found mainly in a range from $-90^{\circ}$ to $90^{\circ}$, whilst $e$ took values from a wide range between 0.002 for $|\omega| \approx 0^{\circ}$ and 0.314 for $|\omega|$ approaching $90^{\circ}$ (these values are given for 99.7\% confidence) with the best-fitting value equal to 0.039.

Because of those weak constraints, we used the circular orbit approximation for WASP-62~b in our analysis. We note, however, that further studies of the WASP-62 system would benefit much from additional precise RV measurements.  

\subsection{Transit modeling}

Photometric transit signatures were modelled with the Transit Analysis Package (TAP, Gazak \etal 2012). Individual transit light curves were extracted from the TESS photometric time series with a time window of $\pm2.5$ times transit duration relative to a transit mid-point predicted by a trial ephemeris. Those out-of-transit measurements were used to account for possible photometric trends in the time domain. TAP was modified to be capable to model such trends with a second-order polynomial. The light curves were inspected by eye and only complete transits were qualified for further analysis. Numbers of qualified transits in individual sectors are listed in Table~2.

To characterise transit geometry, TAP uses the orbital inclination $i_{\rm{orb}}$, the semi-major axis scaled in star radii $a/R_{\star}$, and the ratio of planet to star radii $R_{\rm{p}}/R_{\star}$. The limb darkening (LD) effect is rendered with the quadratic law (Kopal 1950) which is characterised by the linear $u_{\rm 1,TESS}$ and quadratic $u_{\rm 2,TESS}$ coefficients. Those 5 parameters were linked together for all light curves. Mid-transit times $T_{\rm{mid}}$ and trends, in turn, were determined for each transit separately. The orbital period $P_{\rm{orb}}$ was fixed and its value was taken from the transit-timing analysis (Sect.~4.4). 

The LC data were found to be not precise enough to determine the LD coefficient in a reliable way. Thus, these parameters were allowed to vary around the theoretical predictions under the Gaussian penalty with a conservative value of 0.1. The values for the Cousins \textit{R} and \textit{I} bands and the Sloan Digital Sky Survey \textit{z} band were bi-linearly interpolated from tables of Claret \& Bloemen (2011), and then averaged to approximate the TESS bandpass.

For KELT-24~b, the orbital eccentricity and argument of periastron were allowed to vary under the Gaussian penalty with the values refined in Sect.~4.1. For remaining targets, circular orbits were assumed. 

The parameters of the best-fitting models and their uncertainties were determined from the marginalised posteriori probability distributions produced from 10 MCMC chains (\ie the median value, and 15.9 and 84.1 percentiles). The random walk process was governed by the Metropolis-Hastings algorithm and a Gibbs sampler. The correlated noise was modelled with the wavelet-based technique (Carter \& Winn 2009). Each chain was $10^5$ steps long, and the first 10\% of trials were rejected to minimise the influence of the initial values. The best-fitting models are plotted in Fig.~1, and their parameters are collected in Table~4. For ease of comparison, we also list determinations reported in the past studies. The individual mid-transit times are given in Table~5 in Sect.~4.4.

\begin{figure}[thb]
\begin{center}
\includegraphics[width=1.0\textwidth]{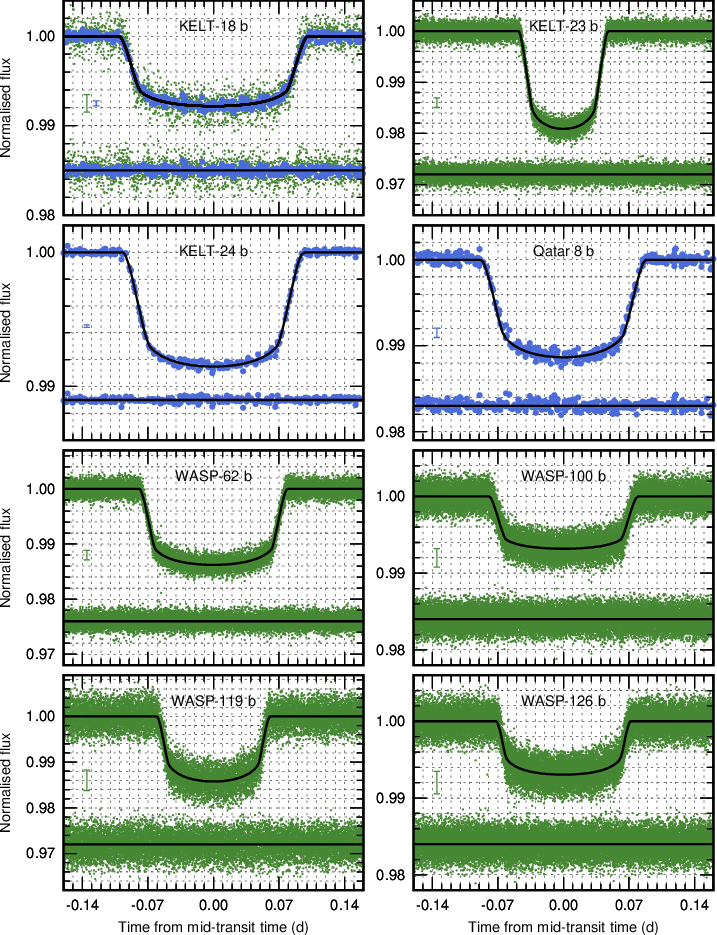}
\end{center}
\FigCap{Phase folded transit light curves with the best-fitting models. The SC and LC measurements are marked with small green and big blue dots, respectively. Typical (median) photometric errors of the individual measurements are shown with error bars in the middle-left side of each panel. KELT-18 was observed with both cadences. The residuals are plotted below each light curve.}
\end{figure}

\MakeTable{ccclll}{12.5cm}{System parameters from transit light curve modeling and from the literature.}
{\hline
$R_{\rm{p}}/R_{\star}$ & $a/R_{\star}$ & $i_{\rm{orb}}$ $(^{\circ})$ & $u_{\rm 1,TESS}$ & $u_{\rm 2,TESS}$ & Reference\\
\hline
\multicolumn{6}{c}{KELT-18 b}  \\
$0.0845^{+0.0005}_{-0.0005}$    & $4.36^{+0.11}_{-0.09}$ & $82.90^{+0.62}_{-0.54}$ & $0.19^{\rm a)}$ & $0.33^{\rm a)}$ & this paper\\
$0.08462^{+0.00091}_{-0.00091}$ & $5.138^{+0.038}_{-0.078}$ & $88.9^{+0.8}_{-1.2}$ & $-$ & $-$ & McLeod \etal (2017)\\
\multicolumn{6}{c}{~}  \\
\multicolumn{6}{c}{KELT-23 b}  \\
$0.1320^{+0.0006}_{-0.0006}$ & $7.556^{+0.041}_{-0.045}$ & $85.96^{+0.11}_{-0.10}$ & $0.30^{+0.07}_{-0.07}$ & $0.23^{+0.12}_{-0.12}$ & this paper\\
                             &                           &                         & $0.27^{\rm b)}$        & $0.29^{\rm b)}$ & \\
$0.1365^{+0.0011}_{-0.0011}$ & $7.13^{+0.17}_{-0.16}$ & $85.37^{+0.31}_{-0.30}$ & $-$ & $-$ & Johns \etal (2019)\\
\multicolumn{6}{c}{~}  \\
\multicolumn{6}{c}{KELT-24 b}  \\
$0.0901^{+0.0003}_{-0.0004}$ & $7.89^{+0.14}_{-0.12}$ & $85.08^{+0.20}_{-0.17}$ & $0.21^{\rm a}$ & $0.32^{\rm a}$ & this paper\\
$0.08677^{+0.00071}_{-0.00070}$ & $9.95^{+0.17}_{-0.18}$ & $89.17^{+0.59}_{-0.75}$ & $-$ & $-$ & Rodriguez \etal (2019)\\
$0.092^{+0.003}_{-0.003}$ & $9.5^{+0.2}_{-0.2}$ & $87.6^{+1.0}_{-0.8}$ & $-$ & $-$ & Hjorth \etal (2019)\\
\multicolumn{6}{c}{~}  \\
\multicolumn{6}{c}{Qatar-8 b}  \\
$0.1014^{+0.0031}_{-0.0026}$ & $6.30^{+0.44}_{-0.36}$ & $84.6^{+1.3}_{-1.1}$ & $0.31^{\rm a)}$ & $0.27^{\rm a)}$ & this paper\\
$0.1005^{+0.0008}_{-0.0008}$ & $7.76^{+0.1}_{-0.1}$ & $89.3^{+0.7}_{-0.7}$ & $-$ & $-$ & Alsubai \etal (2019)\\
\multicolumn{6}{c}{~}  \\
\multicolumn{6}{c}{WASP-62 b}  \\
$0.1105^{+0.0003}_{-0.0003}$ & $9.684^{+0.060}_{-0.061}$ & $88.57^{+0.18}_{-0.16}$ & $0.280^{+0.026}_{-0.023}$ & $0.170^{+0.046}_{-0.052}$ & this paper\\
                             &                           &                         & $0.23^{\rm b)}$           & $0.31^{\rm b)}$ & \\
$0.1109^{+0.0009}_{-0.0009}$ & $9.5^{+0.4}_{-0.4}$ & $88.3^{+0.9}_{-0.6}$ & $-$ & $-$ & Hellier \etal (2012)\\
\multicolumn{6}{c}{~}  \\
\multicolumn{6}{c}{WASP-100 b}  \\
$0.08014^{+0.00040}_{-0.00035}$ & $5.234^{+0.061}_{-0.073}$ & $83.49^{+0.24}_{-0.29}$ & $0.25^{+0.09}_{-0.09}$ & $0.16^{+0.13}_{-0.13}$ & this paper\\
                                &                           &                         & $0.18^{\rm b)}$        & $0.32^{\rm b)}$ & \\
$0.087^{+0.003}_{-0.003}$ & $4.9^{+0.8}_{-0.8}$ & $82.6^{+2.6}_{-1.7}$ & $-$ & $-$ & Hellier \etal (2014)\\
\multicolumn{6}{c}{~}  \\
\multicolumn{6}{c}{WASP-119 b}  \\
$0.1103^{+0.0010}_{-0.0008}$ & $7.26^{+0.10}_{-0.15}$ & $88.6^{+0.9}_{-0.9}$ & $0.37^{+0.08}_{-0.08}$ & $0.17^{+0.17}_{-0.17}$ & this paper\\
                                &                           &                & $0.33^{\rm b)}$        & $0.27^{\rm b)}$ & \\
$0.1145^{+0.0035}_{-0.0035}$ & $6.3^{+0.6}_{-0.6}$ & $85^{+2}_{-2}$ & $-$ & $-$ & Maxted \etal (2016)\\
\multicolumn{6}{c}{~}  \\
\multicolumn{6}{c}{WASP-126 b}  \\
$0.07712^{+0.00063}_{-0.00047}$ & $7.80^{+0.11}_{-0.20}$ & $88.7^{+0.9}_{-0.9}$ & $0.32^{+0.07}_{-0.07}$ & $0.25^{+0.12}_{-0.13}$ & this paper\\
                                &                        &                      & $0.30^{\rm b)}$        & $0.28^{\rm b)}$ & \\
$0.0781^{+0.0013}_{-0.0013}$ & $7.63^{+0.64}_{-0.23}$ & $87.9^{+1.5}_{-1.5}$ & $-$ & $-$ & Maxted \etal (2016)\\
$0.076^{+0.003}_{-0.002}$ & $6.0^{+1.2}_{-1.3}$ & $87.0^{+2.1}_{-4.0}$ & $-$ & $-$ & Feinstein \etal (2019)\\
$0.0780^{+0.0002}_{-0.0002}$ & $7.89^{+0.04}_{-0.04}$ & $89.51^{+0.44}_{-0.44}$ & $-$ & $-$ & Pearson (2019)\\
\hline
\multicolumn{6}{l}{$^{\rm a)}$ Value interpolated from Claret \& Bloemen (2011), varied under the Gaussian penalty.}  \\
\multicolumn{6}{l}{$^{\rm b)}$ Theoretical predictions from Claret \& Bloemen (2011), given for comparison purposes.}  \\
}

For 5 systems observed in SC the mode, \ie KELT-23, WASP-62, WASP-100, WASP-119, and WASP-126, the LD coefficients were  kept free. In Table 4, there are also the theoretical values listed for comparison purposes. For KELT-23, WASP-119, and WASP-126, our values agree with the theoretical ones well within a $1\sigma$ range. For WASP-100, the values of $u_{\rm 1,TESS}$ are consistent within $0.8\sigma$, and $u_{\rm 2,TESS}$ deviates by $1.2\sigma$. For WASP-62, the discrepancy reached $2.2\sigma$ for $u_{\rm 1,TESS}$ and $3.0\sigma$ for $u_{\rm 2,TESS}$.

\subsection{Transit timing}

The set of mid-transit times obtained from TESS photometry was enriched with mid-transit times determined from the ground-based follow-up observations that are available in the literature. We used 11 light curves from McLeod \etal (2017) recorded for 7 transits of KELT-18~b (2 events were observed simultaneously with 3 instruments), and the new light curve reported in this study (Sect.~3.2). We used 11 light curves from Johns \etal (2019) for 7 transits of KELT-23~b (2 events were observed simultaneously with 2 instruments and one  with 3), a phase-folded MASCARA light curve from Hjorth \etal (2019), and 9 light curves for 5 transits of KELT-24~b (4 events observed simultaneously with 2 instruments), 3 light curves for 2 transits of Qatar-8~b from Alsubai \etal (2019), 2 transit light curves for WASP-62~b from Hellier \etal (2012), 3 transit light curves for WASP-100~b from Hellier \etal (2014), 1 transit light curve for WASP-119~b from Maxted \etal (2016), and 5 transit light curves for WASP-126~b from Maxted \etal (2016). 

The mid-transit times were derived by fitting the template transit models that were obtained in Sect.~4.3. The model parameters $i_{\rm{orb}}$ and $a/R_{\star}$ were allowed to vary around the best-fitting values under the Gaussian penalties determined by the parameters' uncertainties. The ratio of planet to star radii $R_{\rm{p}}/R_{\star}$ was linked for all light curves and was allowed vary freely to compensate a possible third-light contamination in the template transit model. The linear and quadratic LD coefficients were bi-linearly interpolated from tables of Claret \& Bloemen (2011) and allowed to vary under Gaussian priors of a conservative width of 0.1. The values of $T_{\rm{mid}}$ and the coefficients of the second-order polynomial that accounts for a possible trend in the time domain were kept free. For the light curves acquired for the same transit, the mid-transit times $T_{\rm{mid}}$ were linked together.

The best-fitting models and their uncertainties were found with TAP following the procedure adopted from Sect.~4.3. The mid-transit times are listed in Table~5. Our ground-based light curve for KELT-18 is plotted in Fig.~2. 

\MakeTable{ l c c c c l}{12.5cm}{Transit mid-points for the examined planets.}
{\hline
Planet      & $E$ & $T_{\rm mid}$ (${\rm BJD_{TDB}}$) & $+\sigma$ (d) & $-\sigma$ (d) & Data source\\
\hline
KELT-18 b   & $-17$ & $2457493.704568$ & $0.000834$ & $0.000857$ & McLeod \etal (2017)\\
KELT-18 b   & $-16$ & $2457496.578632$ & $0.002818$ & $0.002669$ & McLeod \etal (2017)\\
KELT-18 b   & $-1$  & $2457539.655587$ & $0.003451$ & $0.003377$ & McLeod \etal (2017)\\
\hline
\multicolumn{6}{l}{This table is available in its entirety in a machine-readable form at CDS. A portion is }  \\
\multicolumn{6}{l}{shown here for guidance regarding its form and content.}  \\
}

\begin{figure}[thb]
\begin{center}
\includegraphics[width=0.67\textwidth]{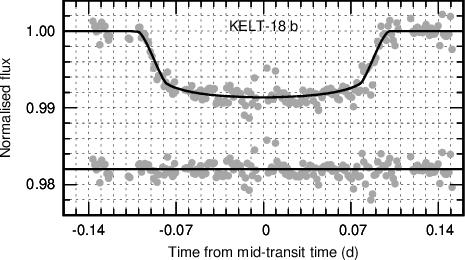}
\end{center}
\FigCap{Ground-based transit light curve for KELT-18~b, acquired on 2017 Feb 27 with a 60 cm telescope. The original measurements were binned with two-minute intervals. The photometric noise is 1.0 ppth of flux per minute of observation. The residuals are shown below.}
\end{figure}

The sets of mid-transit times were used to refine transit ephemerides in a linear form
\begin{equation}
     T_{\rm mid }(E) = T_0 + P_{\rm orb} \cdot E \, , \;
\end{equation}
where $E$ is a transit number counted from a reference epoch $T_0$, taken from a discovery paper\footnote{For KELT-23~b, which was discovered independently by Rodriguez \etal (2019) and Hjorth \etal (2019), transit counting begins from the earliest epoch, \ie given in the latter paper.}. The best-fitting  parameters and their $1\sigma$ uncertainties were determined from posterior probability distributions generated with the MCMC algorithm running 100 chains, $10^4$ steps long each with the first 1000 trials discarded. They are collected in Table~6. The timing residuals against the refined ephemerides are presented in Fig.~3 and 4.  

\MakeTable{ l c c }{12.5cm}{Transit ephemeris elements for the investigated planets.}
{\hline
Planet      & $T_0$ (${\rm BJD_{TDB}}$) & $P_{\rm orb}$ (d)\\
\hline
KELT-18 b   & $2457542.52463 \pm 0.00057$ & $2.8716992 \pm 0.0000013$ \\
KELT-23 b   & $2458140.38698 \pm 0.00020$ & $2.25528783 \pm 0.00000068$ \\
KELT-24 b   & $2458268.45459 \pm 0.00087$ & $5.5514918 \pm 0.0000086$ \\
Qatar-8 b   & $2458210.8392  \pm 0.0010$  & $3.7146428 \pm 0.0000060$ \\
WASP-62 b   & $2455855.39316 \pm 0.00038$ & $4.41193858 \pm 0.00000064$ \\
WASP-100 b  & $2456272.3406  \pm 0.0010 $ & $2.8493819 \pm 0.0000013$ \\
WASP-119 b  & $2456537.5493  \pm 0.0017 $ & $2.4998048 \pm 0.0000023$ \\
WASP-126 b  & $2456890.32004 \pm 0.00061$ & $3.2887884 \pm 0.0000013$ \\
\hline
}

\begin{figure}[thb]
\begin{center}
\includegraphics[width=1.0\textwidth]{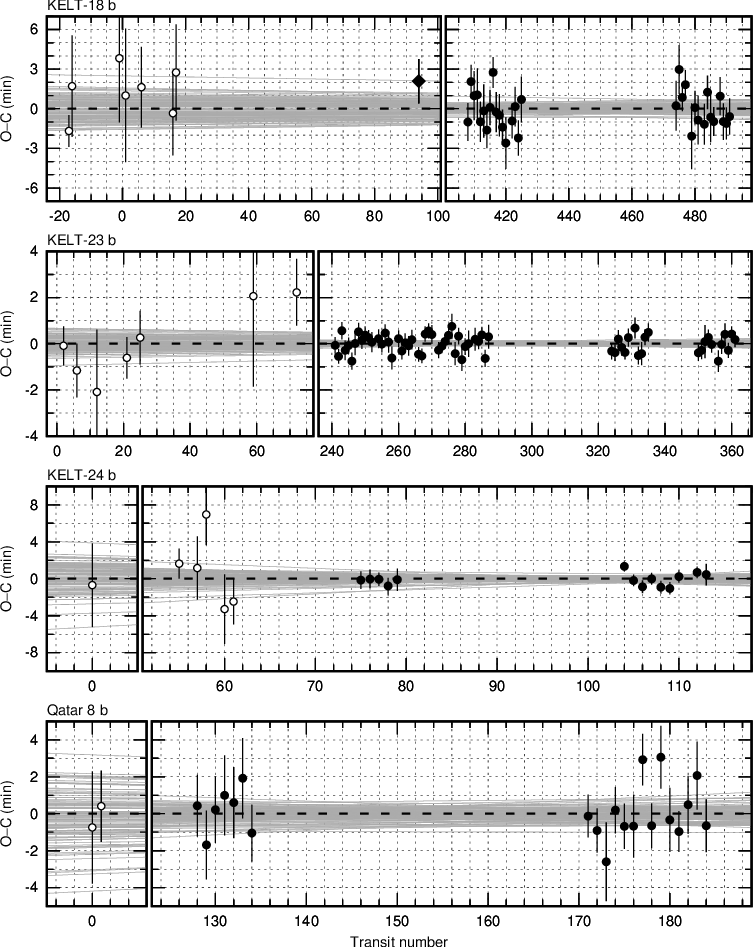}
\end{center}
\FigCap{Transit-timing residuals against the refined linear ephemerides for KELT-18~b, KELT-23~b, KELT-24~b, and Qatar-8~b. The determinations from TESS photometry are marked with dots, and the re-determined literature values are plotted with open circles. For KELT-18~b, a diamond symbol places the result from the ground-base light curve reported in this paper. Dashed lines mark zero value. The ephemeris uncertainties are illustrated by grey lines that are drawn for 100 sets of parameters, chosen from the most likely Markov chains.}
\end{figure}

\begin{figure}[thb]
\begin{center}
\includegraphics[width=1.0\textwidth]{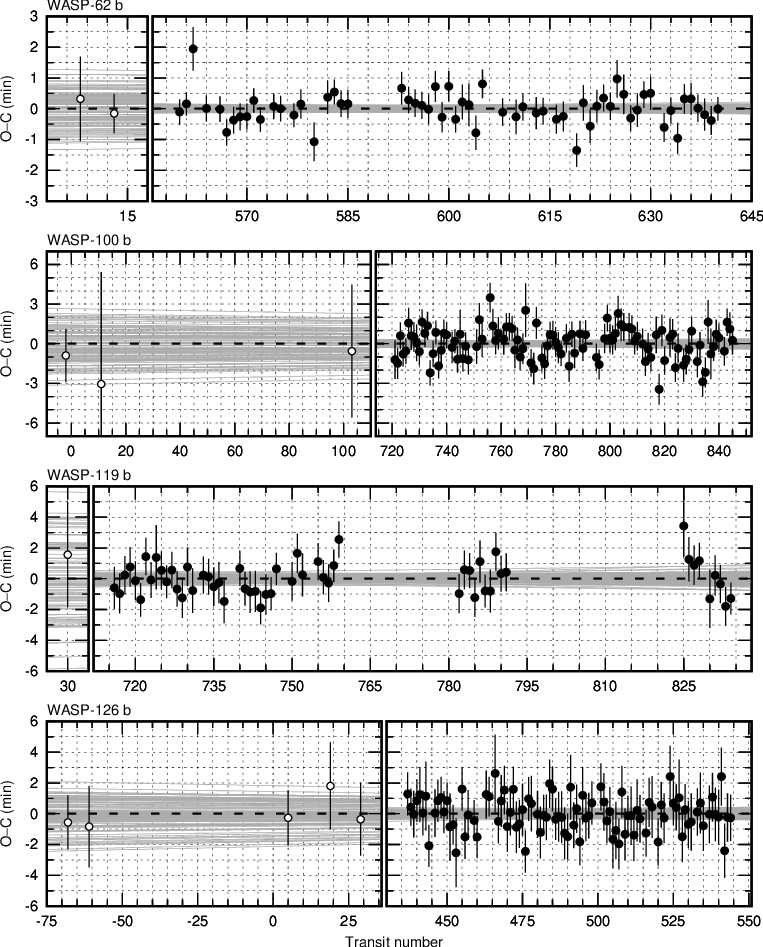}
\end{center}
\FigCap{The same as Fig.~3 but for WASP-62~b, WASP-100~b, WASP-119~b, and WASP-126~b.}
\end{figure}

The timing residuals were searched for long-term trends due to systematic shortening or lengthening of the orbital period. A trial quadratic ephemeris in a form   
\begin{equation}
  T_{\rm{mid}}= T_0 + P_{\rm{orb}} \cdot E + \frac{1}{2} \frac{{\rm d} P_{\rm{orb}}}{{\rm d} E} \cdot E^2 \, , \;
\end{equation}
where ${{\rm d} P_{\rm{orb}}}/{{\rm d} E}$ is the change in the orbital period between succeeding transits, was fitted to check the significance of the quadratic term. For all planets of our sample, the values of ${{\rm d} P_{\rm{orb}}}/{{\rm d} E}$ are consistent within 0 well within $1 \sigma$ range and the linear ephemerides are favoured over their quadratic analogues by the Bayesian information criterion. This result shows that the orbital periods of those planets are stable in long time scales.

The timing residuals were searched for periodic signals employing the analysis of variance algorithm (AoV, Schwarzenberg-Czerny 1996), which was based on a series of 3 Szeg{\"o} orthogonal polynomials. This approach was an equivalent to usage of 1 harmonic, resulting in Generalised Lomb-Scargle periodograms. A periodic TTV signal was sought in a range between 1 and 1000 epochs with a resolution of $5 \times 10^{-5}$ epoch$^{-1}$ in a frequency domain. The false alarm probabilities (FAPs) were empirically determined with the bootstrap resampling method. The $O-C$ values were randomly permuted at the original observing epochs. The FAP levels were determined from the posterior distributions of the periodogram power of the strongest peaks in $10^5$ trials. The results are illustrated in Fig.~5. For all planets in our sample, the strongest power peaks in periodograms were found to be statistically insignificant. 

\begin{figure}[thb]
\begin{center}
\includegraphics[width=1.0\textwidth]{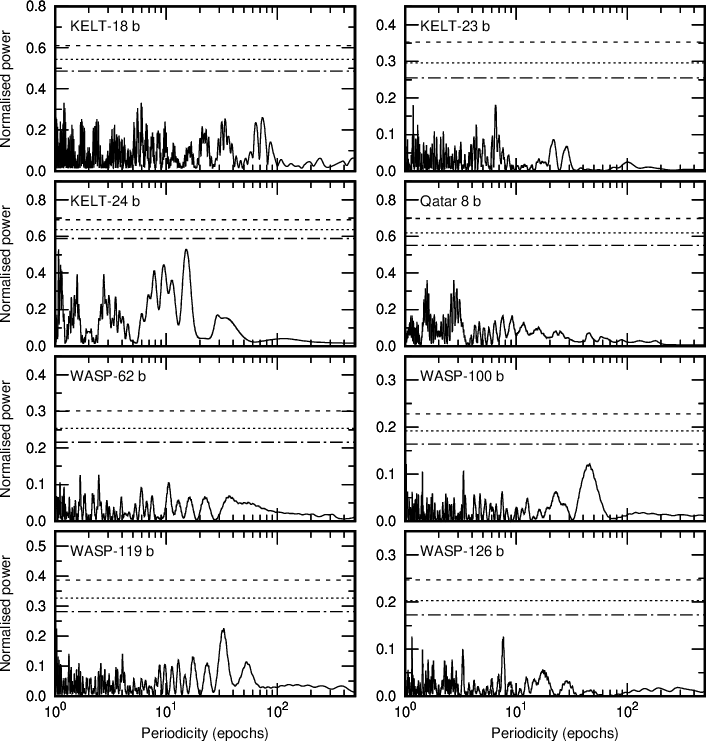}
\end{center}
\FigCap{AoV periodograms for the timing residuals against the linear ephemerides, generated for the planets of our sample. The dashed lines show the empirical FAP levels of 5\%, 1\%, and 0.1\% (from the bottom up).}
\end{figure}

\subsection{Search for additional transiting planets}

The photometric time series were searched for transit signatures, which could be produced by additional transiting planets, using the AoV method optimised for the detection of box-like periodic signals (AoVtr, Schwarzenberg-Czerny \& Beaulieu 2006). In the original light curves, the transits and occultations with a margin of $3 \sigma$ of the transit duration were masked out. The tested periods were in a range 0.5-100 days with a resolution in frequency equal to $2.5 \times 10^{-4}$ day$^{-1}$. The algorithm uses light-curve phase folding and binning to test a negative-pulse model with a minimum associated to a transit event. This method is known to be sensitive to a number of bins and a signal period considered. After a series of trial runs with artificial transit signals injected (as described below), we set the bin number to 100 as an optimised resolution for the considered period range. The FAP levels were determined empirically with the bootstrap method (Sect.~4.4) for $10^4$ resampled datasets.  

The results are illustrated in Fig.~6. For six systems, no statistically significant signals are detected. For WASP-100 and WASP-119, the strongest peaks approach a FAP level of 5\%. However, a careful visual inspection of phase-folded light curves revealed no sign of a real transit-like shape. 

\begin{figure}[thb]
\begin{center}
\includegraphics[width=1.0\textwidth]{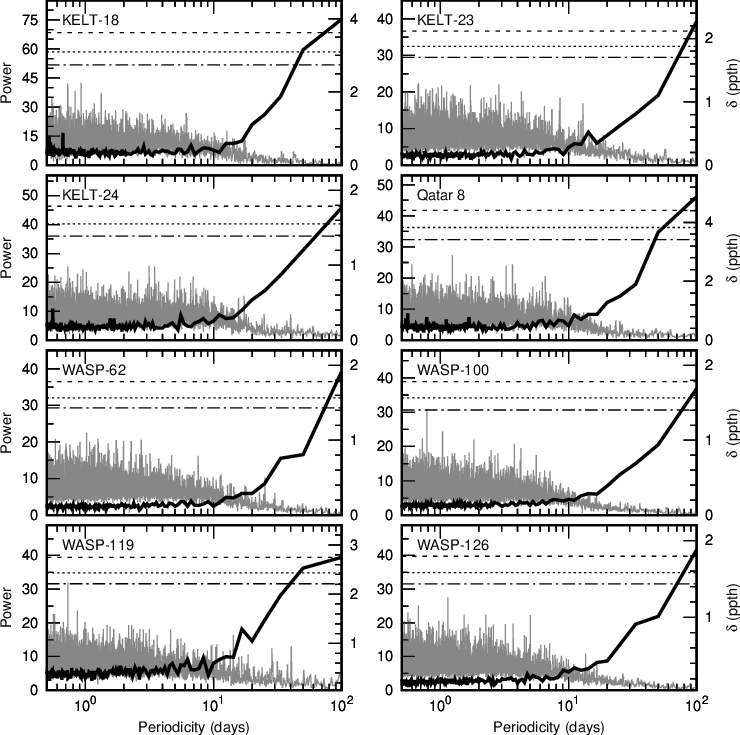}
\end{center}
\FigCap{AoVtr periodograms (on left axes) and upper constraints on depths of transits that avoid detection (right axes) in the examined systems. The power spectra are drawn with grey lines. As in Fig.~5, the dashed lines mark the empirical FAP levels of 5\%, 1\%, and 0.1\% (from the bottom up). The empirical constraints on the transit depths are plotted with black bold lines.}
\end{figure}

To determine a threshold for transits that are too shallow to be detected, artificial transit signals were injected into the original photometric time series. Since this detection limit is expected to depend not only on the quality of photometry but also on a number of transits observed, and hence on an orbital period of an additional planet, trial periods from 0.5 to 100 days with a step in frequency equal to $10^{-2}$ day$^{-1}$ were considered. The synthetic transits were approximated with a box-shape signal of a duration of 2 hours. For each period, the algorithm injected transit signals with depths starting from 0.01 ppth. Then, the depths were iteratively increased with a step of 0.01 ppth. The procedure was stopped when the AoVtr power peak associated to a trial period reached a FAP level of 0.1\%. A series of tests showed that such a transit signal could be unquestionably identified by visual inspection of the phase-folded light curve.

The detection limits for the individual systems are plotted also in Fig.~6. The transit method was found to be the most sensitive for orbital periods shorter than $\approx10$ days, reaching from 0.45 ppth for Qatar-8 to 0.12 ppth for WASP-62. Those quantities were converted into planetary radii using host stars' radii published in the discovery papers. Following the classification proposed by Buchhave \etal (2014), TESS photometry allowed us to probe down to the regime of terrestrial-like planets ($<1.7$ $R_{\oplus}$) for KELT-23 and WASP-62 or mini-Neptunes ($\approx 1.7-3.9$ $R_{\oplus}$) for KELT-24, Qatar-8, WASP-100, WASP-119, and WASP-126. For KELT-18, the detection threshold is at the transition between mini-Neptunes and ice or gas giant planets ($\approx 3.9$ $R_{\oplus}$). The individual results are collected in Table~7. 

\MakeTable{ l c c c}{12.5cm}{Summary of transit detection sensitivity.}
{\hline
System   & $\delta$ (ppth) & $R_{\star}$ $(R_{\odot})$ & $R_{\rm lim}$ $(R_{\oplus})$\\
\hline
KELT-18  & 0.35 & $1.908^{+0.042}_{-0.035}$ & $3.90^{+0.09}_{-0.07}$ \\
KELT-23  & 0.16 & $0.996 \pm 0.015$ & $1.38 \pm 0.02$ \\
KELT-24  & 0.18 & $1.506 \pm 0.022$ & $2.21 \pm 0.03$ \\
Qatar-8  & 0.45 & $1.32 \pm 0.02$ & $3.05 \pm 0.05$ \\
WASP-62  & 0.12 & $1.28 \pm 0.05$ & $1.53 \pm 0.06$ \\
WASP-100 & 0.14 & $2.0 \pm 0.3$ & $2.58 \pm 0.39$ \\
WASP-119 & 0.40 & $1.2 \pm 0.1$ & $2.62 \pm 0.22$ \\
WASP-126 & 0.17 & $1.27^{+0.10}_{-0.05}$ & $1.81^{+0.14}_{-0.07}$ \\
\hline
\multicolumn{4}{l}{$\delta$ is the median of the limiting transit depths for the}  \\
\multicolumn{4}{l}{orbital periods <10 days. The values of $\delta$ were  }  \\
\multicolumn{4}{l}{converted into planetary radii $R_{\rm lim}$ using stellar radii $R_{\star}$ }  \\
\multicolumn{4}{l}{which are taken from the discovery papers. }  \\
}


\section{Discussion}

The radii ratio $R_{\rm{p}}/R_{\star}$  were found to be consistent within $1\sigma$ with the literature values for KELT-18~b, Qatar-8~b, and WASP-62~b. For KELT-24~b, our determination of $R_{\rm{p}}/R_{\star}$ agrees within $0.6\sigma$ with the result of Hjorth \etal (2019), but is $4.1\sigma$ greater than the value reported by Rodriguez \etal (2019). Our determination of $R_{\rm{p}}/R_{\star}$ for WASP-126~b is consistent within $1\sigma$ with results reported by both Maxted \etal (2016) and Feinstein \etal (2019), but smaller by $1.3\sigma$ as compared to the value obtained by Pearson (2019).  
For KELT-23~b, $R_{\rm{p}}/R_{\star}$ was found to be smaller by $3.6\sigma$ than the result of Johns \etal (2019). There are no other bright stars in the aperture used and shallower transits are unlikely to be caused by third light contamination\footnote{With a TESS pixel scale of 21'' and with star's pixel response function occupying typically an aperture region of $4 \times 4$ pixels, the probability of blending is relatively high.}. Our determinations of $R_{\rm{p}}/R_{\star}$ for WASP-100~b and WASP-119~b are smaller by 2.3 and 1.2$\sigma$, respectively, than the values reported in the discovery papers. Since there is a $\approx 3$~mag fainter star in an angular distance of 28'' from WASP-100's position and 2 stars fainter by $\approx 3.5$~mag within 36'' from WASP-119, these shallower transits are attributed to the third light contamination.

Our determinations of $a/R_{\star}$ and $i_{\rm{orb}}$ and those from the literature agree well within $1\sigma$ with each other for WASP-62~b and WASP-100~b, and are up to order of magnitude more precise. For WASP-126, they are consistent with values brought by both Maxted \etal (2016) and Pearson (2019), whilst the value of $a/R_{\star}$ from Feinstein \etal (2019) is smaller by 1.5$\sigma$. Our values of $a/R_{\star}$ and $i_{\rm{orb}}$ differ by $1.6-2.4\sigma$ from the previous determinations for KELT-23~b and WASP-119~b. For Qatar-8~b, the discrepancy is at a 3.2$\sigma$ level. Our solution favours a moderate value of $i_{\rm{orb}}$ and smaller $a/R_{\star}$. Significant discrepancies were found for KELT-18~b and KELT-24~b. In the former system, $i_{\rm{orb}} = 82.90^{+0.62}_{-0.54}$ degrees appears to be 4.4$\sigma$ lower than the value of $88.9^{+0.8}_{-1.2}$ obtained by McLeod \etal (2017). Consequently, $a/R_{\star} = 4.36^{+0.11}_{-0.09}$ differs by $5.8\sigma$ from the previously reported $5.138^{+0.038}_{-0.078}$. The similar conclusion can be reached for KELT-24~b. The $i_{\rm{orb}} = 85.08^{+0.20}_{-0.17}$ degrees is smaller by 5.3 and 3.1$\sigma$ from results of Rodriguez \etal (2019) and Hjorth \etal (2019), respectively.  The value of $a/R_{\star} = 7.89^{+0.14}_{-0.12}$, in turn, was found to be smaller by 9.0 and 6.6$\sigma$. 

The orbit of both planets, KELT-24~b and WASP-62~b, have non-zero eccentricities that could be excited and sustained by non-Keplerian planet-planet interactions. Our search for additional planets in their systems has brought, however, the negative results. Although the age of the KELT-24 system is rather poorly constrained, ranging from $0.78^{+0.61}_{-0.42}$ Gyr (Rodriguez \etal 2019) to $2.8^{+0.5}_{-0.6}$ Gyr (Hjorth \etal 2019), it is comparable or even shorter than a circularisation time scale for KELT-24~b. Using Eq.\ (3) of Adams \& Laughlin (2006) and assuming that a value of a planetary tidal quality factor is $10^5 - 10^6$, this time scale is in a range of $\approx 1.2-12$ Gyr for the system parameters from Rodriguez \etal (2019) and $\approx 0.7-6.5$ Gyr for the parameters from Hjorth \etal (2019). The age of the WASP-62 system can be constrained either to $\approx 0.5$ Gyr from lithium absorption in star's spectrum and or to $0.7^{+0.4}_{-0.3}$ Gyr from gyrochronology (Hellier \etal 2012). For a conservative value of the planetary tidal quality factor of $10^6$, WASP-62~b's circularisation time-scale is $\approx 0.3$ Gyr, \ie of the same order as the system's age. The non-zero eccentricities of both planets can be thus plausibly explained as the remnants of high-eccentricity migration, and the circulation process is still ongoing.

Our analysis of transit times for WASP-126~b does not confirm the presence of the variations that were reported by Pearson (2019). Our transit times from sectors 1--3 show no periodic modulation, though for some of the most scattered data points the residuals are qualitatively similar. The additional data from sectors 4--13 clearly strengthen the conclusion that WASP-126~b is an unperturbed planet.

\section{Conclusions}

Our analysis has resulted in a substantial improvement of the precision of determinations of transit parameters in 6 systems: KELT-23, KELT-24 (including orbital parameters), Qatar-8, WASP-62, WASP-100, and WASP-119. For 3 systems: KELT-18, KELT-24, and Qatar-8, their redetermined parameters were found to differ significantly or moderately from values reported in the literature. The transit timing analysis and direct transit search have revealed no sign of nearby planetary companions in all of the 8 examined systems, including WASP-126. This finding is in line with the observation that hot Jupiters in compact planetary systems are rare, and speaks in favour of the high-eccentricity migration as a mechanism shaping their orbital architectures.


\Acknow{We thank the anonymous referee for remarks that improved this paper. We are grateful to Dr.~Coel Hellier, Dr.~Daniel Johns, and Dr.~Zlatan Tsvetanov for sharing the light curves with us. We are also grateful to Prof.~Alex Schwarzenberg-Czerny for developing the AoV package. GM acknowledges the financial support from the National Science Centre, Poland through grant no. 2016/23/B/ST9/00579. This paper includes data collected with the TESS mission, obtained from the MAST data archive at the Space Telescope Science Institute (STScI). Funding for the TESS mission is provided by the NASA Explorer Program. STScI is operated by the Association of Universities for Research in Astronomy, Inc., under NASA contract NAS 5-26555. This research made use of Lightkurve, a Python package for Kepler and TESS data analysis (Lightkurve Collaboration, 2018). This research has made use of the SIMBAD database and the VizieR catalogue access tool, operated at CDS, Strasbourg, France, and NASA's Astrophysics Data System Bibliographic Services.}


\end{document}